\documentclass{article}

\usepackage{arxiv}

\usepackage[utf8]{inputenc} 
\usepackage[T1]{fontenc}    
\usepackage{hyperref}       
\usepackage{url}            
\usepackage{booktabs}       
\usepackage{amsfonts}       
\usepackage{nicefrac}       
\usepackage{microtype}      
\usepackage{lipsum}
\usepackage{graphicx}
\graphicspath{ {./images/} }

\usepackage{tikz}
\usetikzlibrary{positioning}
\usetikzlibrary{arrows.meta, positioning}

\title{Responsible Blockchain: STEADI Principles and the Actor-Network Theory-based Development Methodology (ANT-RDM)}

\author{
Yibai Li \\
University of Scranton \\
yibai.li@scranton.edu \\
   \And
Ahmed Gomaa \\
University of Scranton \\
ahmed.gomaa@scranton.edu \\
  \And
Xiaobing Li \\
University of Scranton \\
xiaobing.li@scranton.edu \\
}

\begin{document}
\maketitle

\begin{abstract}
This paper provides a comprehensive analysis of the challenges and controversies associated with blockchain technology. It identifies technical challenges such as scalability, security, privacy, and interoperability, as well as business and adoption challenges, and the social, economic, ethical, and environmental controversies present in current blockchain systems. We argue that responsible blockchain development is key to overcoming these challenges and achieving mass adoption. This paper defines Responsible Blockchain and introduces the STEADI principles (sustainable, transparent, ethical, adaptive, decentralized, and inclusive) for responsible blockchain development. Additionally, it presents the Actor-Network Theory-based Responsible Development Methodology (ANT-RDM) for blockchains, which includes the steps of problematization, interessement, enrollment, and mobilization.
\end{abstract}


\section{Introduction}
Bitcoin, introduced by Satoshi Nakamoto in 2008, marked the inception of the first peer-to-peer currency \cite{nakamoto2008bitcoin}. Since its inception, many other cryptocurrencies have been introduced and have gained significant traction. As of the end of 2022, it is estimated that the number of cryptocurrency owners grew to 425 million worldwide \cite{cryptocom2022}. Although Nakamoto's original paper mentioned a "chain of blocks," the term "blockchain" itself did not appear in the 2008 white paper. Instead, "blockchain" evolved as a loose umbrella term \cite{narayanan2017bitcoin} within the cryptocurrency community to describe a suite of technologies—including decentralized ledgers, linked timestamping, Merkle trees, consensus mechanisms, and public keys as identities—that underpin Bitcoin. These technologies have been recognized for their potential to revolutionize various sectors such as manufacturing, construction, healthcare, finance, insurance, supply chain, agriculture, academic publishing, energy, resource management, and legal systems \cite{haq2023concept}, overcoming challenges related to information sharing, traceability, and operational efficiency. 

Despite the significant interest and hype about cryptocurrencies and blockchains, more than 10 years since their inception, they have still not become everyday technologies for consumers \cite{alshamsi2022systematic}. This technology still faces many technical challenges, such as scalability, security, privacy, interoperability, and energy consumption, as well as business adoption challenges, social, ethical, environmental, and regulatory controversies. We argue that responsible blockchain development is the key to overcoming these challenges and achieving mass adoption.

\paragraph{Responsible Blockchain Definition}

In this paper, we define responsible blockchain as a socio-technical system that fosters sustainable, transparent, ethical, adaptive, decentralized, and inclusive practices among diverse participants to promote a dynamic equilibrium of interests for its long-term viability. The participants include human actors such as developers, users, regulatory entities, and non-human actors such as computer hardware, software, protocols, policies, and the environment.

This paper will provide a comprehensive analysis of the challenges and controversies of blockchain technology, identify the technical challenges such as scalability, security, privacy, and interoperability, and also the business and adoption challenges, and social, economic, ethical, and environmental controversies within current blockchain systems. The paper will then introduce the STEADI principles (sustainable, transparent, ethical, adaptive, decentralized, and inclusive) aimed at fostering responsible blockchain development, grounded in Actor-Network Theory. Additionally, this paper presents the Actor-Network Theory-based Responsible Development Methodology (ANT-RDM) for blockchain technology, incorporating the stages of problematization, interessement, enrollment, and mobilization.

\section{Challenges of Blockchain Technology}

This section explores the multifaceted challenges associated with blockchain technology, emphasizing technical barriers such as scalability, security, privacy, interoperability, and energy consumption. In addition to technical challenges, this section delves into business and adoption hurdles, including high setup and maintenance costs, slow transaction speeds, regulatory uncertainties, and environmental concerns.

\subsection{Technical Challenges}

This section outlines the significant technical challenges that hinder the widespread adoption of blockchain technology, including scalability, security, privacy, interoperability, and energy consumption.

\subsubsection{Scalability}
Scalability poses a significant obstacle to the widespread adoption of blockchain technology \cite{10.1016/j.jnca.2021.103232}. As the number of users and transactions on a blockchain network increases, its capacity to process these transactions efficiently without compromising security begins to diminish. This challenge is difficult to overcome due to the inherent conflict among three key pillars of blockchain technology: decentralization, security, and scalability. This conflict is often referred to as the "scalability trilemma"  \cite{10.1145/3410699.3413800}. 

Efforts to increase blockchain throughput often result in compromises in decentralization through mechanisms such as increasing block size  \cite{10.1145/3390566.3391673}, implementing sharding  \cite{10.1145/3404397.3404460} or consolidating validation power as seen in the Proof-of-Stake (PoS) consensus mechanism \cite{8752710}. These approaches can pose security risks by potentially creating powerful nodes capable of controlling or manipulating the blockchain.

\subsubsection{Security}
Blockchain, known for its features like immutability and decentralization, plays a crucial role in enhancing the security of various applications and services. However, the security of blockchain itself often gets overlooked. Blockchain can be considered a five-layer architecture, which includes the Hardware/Infrastructure Layer, Data Layer, Network Layer, Protocol (Consensus) Layer, and Application Layer \cite{birje2022review, gomaa2022entrepreneurial}. Each layer of the blockchain architecture has its own set of security issues.

\paragraph{Hardware-Level Attacks} focus on the physical devices and components within the blockchain network. These attacks include Backdoor Trojan attacks, where malicious software is installed on hardware to gain unauthorized access or control \cite{Zhai2022BlockchainAndTime}. Additionally, Side-Channel attacks exploit indirect information from a cryptographic system's physical implementation. Attackers gather secret information by analyzing operational side effects, such as power consumption or timing \cite{saravanan2019optimized}.

\paragraph{Data Layer Attacks} specifically target the data component of blockchain transactions. One example is double spending, which refers to the situation where the same digital currency (or digital asset) is spent more than once \cite{begum2020blockchain}. Another example, the Malleability attack \cite{khan2020simulation}, is a variant of the double spending attack, derived from the malleability of signatures \cite{khan2020simulation}. A Hash collision attack involves finding two different inputs that generate the same hash output. If successful, it could allow attackers to manipulate transactions or create counterfeit data \cite{elsayed2019collision}.

\paragraph{Network Attacks} encompass various strategies to disrupt blockchain operations. Denial-of-Service (DoS) attacks flood the network with excessive data, overwhelming resources and blocking legitimate access \cite{mirkin2020bdos}. In a Sybil attack, an attacker uses multiple fake identities (Sybil nodes) to influence the network and disrupt consensus mechanisms, potentially manipulating transaction validation or launching a 51\% attack \cite{rajab2020feasibility}. Eclipse attacks isolate a node from the network, hindering its ability to receive updates and influencing its consensus decisions \cite{dai2022eclipse}. Lastly, routing attacks manipulate network protocols to redirect traffic through malicious nodes, enabling data interception or injection \cite{sahay2020novel}.

\paragraph{Consensus Attacks} Among various attacks, the 51\% attack is particularly critical. In Proof of Work blockchains, an individual or group gains control of more than 50\% of the network's mining power. This dominance allows them to manipulate the consensus process, enabling actions such as reversing transactions, double-spending coins, or blocking legitimate transactions \cite{aponte2021mining}. Proof of Stake introduces an economic disincentive for such attacks. The assumption is that if a 51\% attack succeeds, the value of the cryptocurrency will fall. Since the attacker holds a large amount of this currency, they would suffer significant losses. However, attackers could engage in short selling the attacked cryptocurrency in another market and profit from the subsequent decrease in value, which follows the chaos and loss of confidence triggered by the attack \cite{lee2020proof}.  Another attack primarily be found in Proof of Work blockchains is selfish mining, where miners withhold their blocks from the network until certain conditions are met, thereby gaining an unfair advantage \cite{bai2019selfish}. 

\paragraph{Smart Contract Attacks} The re-entrancy attack targets vulnerabilities in smart contracts, enabling attackers to repeatedly withdraw funds from a single transaction \cite{alkhalifah2021ethereum}. The Transaction-Ordering Dependence (TOD) attack exploits the manipulation of transaction sequences to gain an unfair advantage, particularly in scenarios where the order of transactions is crucial \cite{sayeed2020smart}. Another significant threat is the front-running attack, where attackers observe pending transactions and execute trades before them, taking advantage of anticipated price movements \cite{momeni2022fairblock}.

\paragraph{Application Layer Attacks} Phishing attacks are a prominent threat where cybercriminals use fake emails and websites to masquerade as legitimate entities. The goal is to deceive users into disclosing sensitive information such as login credentials \cite{andryukhin2019phishing}. Social engineering attacks, another serious concern, exploit human vulnerabilities, manipulating individuals into divulging confidential information or transferring funds to malicious actors \cite{weber2020exploiting}. Furthermore, exchange hacks pose a significant risk to cryptocurrency exchanges by targeting and compromising user accounts, leading to the theft of funds \cite{hong2019survey}.

\subsubsection{Privacy}

Blockchain's transparency and immutability facilitate trust and auditability; however, they also present a paradoxical challenge for data privacy. Despite the common perception that blockchain transactions are anonymous, they are, in fact, pseudonymous \cite{werner2020blockchain}. Blockchain ledgers record transactions publicly, linking them to unique addresses \cite{Zheng2018}. Although these addresses do not directly identify individuals, they can be linked to real-world identities through various means, such as deanonymization attacks \cite{zhang2020deanonymization} and on-chain analysis \cite{lv2020study}. Once recorded, transactions and associated data become permanently etched on the chain, accessible to anyone. This permanence impedes individuals' right to rectification or erasure \cite{Moreno2019Blockchain}.

\subsubsection{Interoperability}
Unlike the seamless flow of information across the internet, different blockchains often operate in isolation, unable to communicate or exchange data effectively \cite{8726757}. Simple operations, such as transferring assets across different platforms, can be very difficult to achieve without trusted custodians like cryptocurrency exchanges, which, on the other hand, can undermine decentralization \cite{kang2022blockchain}. This fragmentation leads to interoperability issues, hindering the true potential of blockchain technology.

One of the primary reasons for these interoperability issues is the heterogeneity of blockchain platforms \cite{cmu212594}. Early blockchain projects, such as Bitcoin, did not prioritize interoperability in their design \cite{9872938}. Now, with thousands of blockchains and cryptocurrencies created, the importance of interoperability is becoming more apparent \cite{10174872}. Each platform boasts unique features, consensus mechanisms, and programming languages, yet there is a lack of standardized protocols for cross-chain communication \cite{OU2022109378}. While solutions like cross-chain bridges exist, they suffer from a lack of universal adoption \cite{10.1145/3573896}. These issues are hindering the growth of a truly interconnected blockchain ecosystem.

Security and privacy considerations also play a significant role in these interoperability challenges \cite{10.1145/3530019.3531345}. Maintaining trust and the immutability of data when transferring between platforms is crucial. Securely verifying the authenticity of data originating from another blockchain requires robust mechanisms that ensure data integrity and prevent malicious manipulation \cite{10150492}. Striking the right balance between security and efficient cross-chain communication remains a complex challenge.

Beyond technical hurdles, the governance models of different block-chains can also create friction \cite{10.3233/IP-190154}. Decentralized governance and independence are core values of blockchain; however, reaching a consensus on how to integrate and manage data exchange across platforms with diverse governance structures can be an arduous and time-consuming process. A case study \cite{8428782} compares the governance mechanisms of Bitcoin and Dash. Dash utilizes the Decentralized Governance By Blockchain (DGB) process \cite{dash2023}, while Bitcoin relies on the Bitcoin Improvement Proposal (BIP) process \cite{cryptoeprint:2013/829}. This difference in models significantly impacted their decision-making speed. Dash could decide on altering the block size in just a few hours, whereas Bitcoin's governance took several years to reach the same decision.

\subsubsection{Energy Consumption}
The energy consumption associated with blockchain and cryptocurrency operations is significant. Globally, the energy usage of blockchain technology is estimated to exceed 100 terawatt-hours (TWh) annually \cite{coroama2021exploring}. To contextualize this, the U.S. Energy Information Administration (EIA) reported that, in 2021, the average American household consumed approximately 10,632 kilowatt-hours (kWh) of electricity per year \cite{epa_greenpower}. Thus, the energy used by blockchain technologies is sufficient to power an estimated 9.4 million U.S. households.

The energy consumption problem is particularly pronounced in blockchain systems that utilize the Proof of Work (PoW) consensus mechanism \cite{alofi2022optimizing} such as Bitcoin. There are three main sources of energy consumption in blockchain operations: data storage, the computation required for PoW, and communication between nodes. An economic threshold analysis \cite{coroama2021exploring} reveals that for a typical PoW blockchain consuming 100 TWh annually, data storage accounts for 50 MWh to 4.25 GWh, which is only 0.00005\% to 0.00425\% of the total consumption.  Communication between nodes uses about 88 MWh (0.000088\%), while a staggering 99.99\% of energy is consumed by the mining process. his process is particularly energy-intensive because it requires network participants (miners) to competitively solve complex cryptographic mathematical puzzles, demanding substantial computational power \cite{pandya2022gpu}. 

The energy consumption is crucial to the security and reliability of the Bitcoin network \cite{treiblmaier2023comprehensive}. However, the sustainability of such high energy usage, particularly from non-renewable sources, becomes increasingly questionable as these blockchains expand. This concern has spurred the exploration of alternative consensus mechanisms, such as Proof of Stake (PoS). A significant transition occurred when Ethereum completed its "Merge" on September 15, 2022, moving from Proof of Work (PoW) to PoS. According to the Ethereum \cite{ethereum_energy_usage}, this shift could reduce its energy consumption by approximately 99.95\%.

\subsection{Business and Adoption Challenges}

In addition to the technical challenges discussed, widespread business adoption also faces hurdles. These include significant setup and maintenance costs, along with the necessity for regular updates \cite{panghal2023adoption}. Additionally, the regulatory landscape for digital assets remains unclear, with differing views including money, property, commodity, and security \cite{goforth2018us} to issue initial coin offerings (ICOs) \cite{cai2019initial}, complicating their integration with existing financial systems \cite{trivedi2021systematic}.

To address these challenges requires a joint effort \cite{toufaily2021framework}. The Blockchain Innovation Adoption Framework reveals that both organizational and individual factors are crucial \cite{upadhyay2020demystifying}. For small and medium businesses, management support, affordability, and regulatory guidance are key to leveraging blockchain \cite{wong2020time}. The adoption of blockchain is influenced by various factors. In supply chains, its benefits and external pressures play a significant role \cite{agi2022blockchain}. Readiness and top management support are essential for successful implementation, as evidenced by research in Ireland \cite{hua2019current}. While blockchain can offer transparency for consumers, the associated costs may deter manufacturers \cite{iyengar2023economics}. Blockchain's adoption can also impact the gray market, influencing manufacturers' pricing strategies and gray marketers' entry based on additional costs and product quality in foreign markets \cite{zhang2023blockchain}. The strategic risks of adopting blockchain encompass business, legal, and technological considerations \cite{stratopoulos2022use}. Initial blockchain implementation in companies is shaped by the novelty of the technology, associated costs, and external scrutiny \cite{guo2021early}.

\section{Controversies Surrounding Blockchain Technology}
Despite the potential of blockchain to transform society by introducing transparency, trust, and immutability, it is not without its detractors and dilemmas. This section analyzes the controversies surrounding blockchain technology, including the social, ethical, environmental, and regulatory controversies it faces. From the digital divide to privacy concerns, and illicit activities, this section highlights the need for responsible approaches in blockchain development and governance.

\subsection{Social and Ethical Controversies}
Blockchain, despite its potential for revolutionizing various industries, also faces a critical hurdle: the \textbf{digital divide}. The problem of the digital divide arises primarily from the uneven distribution and application of this technology across different nations and societies \cite{Gillpatrick_Boğa_Aldanmaz_2022}. Disruptive technologies like blockchain can impact growth, employment, and inequality by creating new markets and business practices and necessitating new product infrastructures and work skills. However, not all societies and economies are equally positioned to adopt these technologies. This reality contributes to the broadening of the digital divide, not just between developed and underdeveloped nations \cite{kshetri2018blockchain} but also between rural and urban populations \cite{igboanusi2020blockchain}, and between genders \cite{DIVAIO2023102517}.

Public blockchains, such as Bitcoin, are known for their immutability and transparency. Each transaction is permanently recorded on an immutable ledger, which is openly accessible. However, this openness can infringe upon user \textbf{privacy} and may conflict with laws like the General Data Protection Regulation \cite{Finck2019}. This regulation, effective since May 2018 \cite{GDPR2016}, includes provisions such as the \textbf{"right to erasure"} or \textbf{"right to be forgotten"} (RtbF), creating significant concerns for users of public blockchain systems.

To align blockchain's immutability with legal requirements such as the Right to be Forgotten (RtbF), the concept of a redactable blockchain has been introduced. For instance, a specific model known as the k-time modifiable and epoch-based redactable blockchain (KERB) allows participants to modify content \cite{9521222}. This model imposes monetary penalties to deter and penalize malicious actions. However, the implementation of such mutable blockchain systems contradicts the original principles of blockchain technology, presenting challenges in maintaining integrity and trust within the blockchain ecosystem. \cite{politou2020delegated} suggest using the InterPlanetary File System (IPFS) protocol to enable the original content provider, or their delegates, to issue an erasure request across all IPFS nodes. Nevertheless, they also recognize that fully enforcing content erasure in a decentralized network like IPFS is a challenging task.

\subsection{Environmental Impact}
The environmental impact of blockchain technology is a pivotal topic of debate. On one hand, critics point to environmental degradation associated with the energy-intensive proof-of-work mining process, which not only increases CO2 emissions but also leads to the rapid obsolescence of mining equipment, and significant e-waste  \cite{todorovic2019unsustainability}. Research \cite{mohsin2023crypto, karatas2023nonlinear, gupta2022nexus} has identified empirical evidence of a causal link between cryptocurrency activity and environmental degradation, suggesting both bidirectional \cite{mohsin2023crypto} and unidirectional \cite{zhang2023role, ramirez2022evaluation} relationships between them. These findings highlight the urgent need for the industry to adopt green technologies and implement fiscal reforms to reduce its ecological footprint, as advocated by \cite{mohsin2023crypto}.

On the other hand, blockchain offers promising solutions for environmental sustainability. It improves sustainability across various sectors by enhancing traceability and transparency, notably through smart contracts \cite{yousefi2022analytical}.  Blockchain disrupts traditional industries, aiding them in achieving the UN's sustainable development goals \cite{hughes2019blockchain}. Additionally, it supports circular economy strategies—such as facilitating markets for second-hand goods— which may significantly reduce environmental impacts by changing the way materials and natural resources are valued and traded \cite{herweijer2018building}. It potentially cuts environmental footprints by up to 53.8\% in these applications  \cite{shou2022integrating}.

\subsection{Potential for Illicit Activities}
The anonymity provided by blockchain technology, particularly in the case of Bitcoin, presents a double-edged sword. While it offers privacy for users, it simultaneously opens avenues for illicit activities \cite{8802640}. The Bitcoin Blockchain, with its distributed and openly accessible ledger, conceals the real-world identities of entities behind pseudonyms, known as addresses. This inherent anonymity in Bitcoin is widely believed to contribute to its utilization in illegal transactions \cite{foley2019sex}, offering a high level of privacy to its users. Still, Supervised Machine Learning may predict, with an average cross-validation accuracy of 80.42\%, the characteristics of entities that have not yet been identified \cite{sun2019regulating}. Techniques such as heuristics and graph analysis have proven effective in unraveling the behaviors of Bitcoin addresses and transactions. These methods enable the identification of potential red flag indicators and the analysis of patterns and typologies associated with illicit behavior \cite{turner2018bitcoin}. The role of crypto asset mixers, like Tornado Cash, highlights the need for a balanced approach, allowing financial market regulators to address illegal activities while enabling honest users to engage with privacy-enhancing protocols \cite{N.S2023TornadoCash, holt2023assessment}. The lack of a universally accepted digital forensic framework for investigating related crimes underscores the challenges faced by law enforcement and regulatory bodies in adapting to the nuances of cryptocurrency-related investigations, complicating the task of maintaining legal and financial order in the digital currency space \cite{park2023forensic}. The misconception of absolute anonymity was further clarified by the case of Coinbase v. U.S. \cite{mark2018court}. The court's decision, in this case, authorized the IRS to acquire user data from Coinbase with defined limitations. As highlighted by \cite{10.1093/ajcl/avac022}, the Coinbase case shows that even with strict theoretical regulations, practical solutions prevail when monitoring millions of transactions to apply the law and preserve users' privacy. The needed balance of anonymity, trust, and the ability to audit to ensure the prevention of illicit activities conducted via cryptocurrencies is still an area of research and policy exploration \cite{akanfe2024blockchain}.

\subsection{Centralization and Governance Issues}
Proof of stake (PoS), and other algorithms offer faster validation times and reduced energy consumption \cite{8752710}. However, they also present a concern: those with more stake hold more power in the validation process. This raises issues about censorship and manipulation of the network by large token holders \cite{8752710}. It creates barriers to entry for smaller participants, deviating from the initial idealism of blockchain being a grassroots movement \cite{Nouyrigat2019}.

Blockchain governance models encompass critical elements like access rights, decision-making power, incentive structures, accountability mechanisms, and conflict resolution schemes \cite{liu2022defining}. Underpinning this dynamic governance is a large number of consensus mechanisms, each crafted for specific contexts \cite{lashkari2021comprehensive}. There are 130 consensus mechanisms \cite{lashkari2021comprehensive}, and the number is growing, across various blockchain platforms, with applications in multiple domains, including improved supply chain management \cite{dutta2020blockchain}, building trust, and improving efficiency with platforms like Blocktivity \cite{nguyen2023blockchain}. They facilitate improved healthcare management with solutions like BurstIQ \cite{HALEEM2021130} and empower decentralized applications through Hyperledger \cite{khan2023critical}. Different implementation techniques, such as Directed Acyclic Graph (DAG)-based approaches, allow for greater agility in Internet of Things applications \cite{cao2019internet}. Furthermore, they enable streamlined real estate transactions \cite{GAO2022} and reduce the cost of international payments \cite{prasad2023will}, while being cautious of causing harm \cite{galanti2023can}. Establishing robust regulations for digital assets is vital to ensure market stability and protect consumers and investors \cite{mateen2023regulation}.

\section{Responsible Blockchain Development Methodology and Principles}

The field of blockchain development is enriched by a variety of methodologies and design principles, as suggested by both the research community \cite{lockl2020toward, rahman2020design} and industry experts \cite{Maksym125126}. However, the predominant focus of these frameworks is on the technical aspects of blockchain technology. This perspective, while crucial, overlooks a fundamental aspect of blockchain: it is not merely an IT artifact characterized solely by its technical features. Rather, blockchain represents a complex social network and ecosystem, intricately woven into society \cite{khan2022graph}. It operates within an environment that encompasses both human and non-human actors, each contributing to and being influenced by the blockchain.

Recognizing this broader context, this paper endeavors to develop an Actor-Network Theory-based Responsible Development Methodology (ANT-RDM) and establish a set of design principles for blockchain development that extend beyond technical considerations. These principles aim to address the ethical and social responsibilities inherent in the creation of blockchain technology, particularly concerning the diverse actors involved in and affected by this environment. By integrating these considerations, the proposed design principles seek to foster a more holistic approach to blockchain development, one that acknowledges and respects the multifaceted nature of the technology and its far-reaching implications in society.

Actor-Network Theory (ANT) serves as the primary theoretical lens guiding our methodology and principles. ANT is a theoretical and methodological approach used in social science, particularly in the fields of science and technology studies, organizational studies, and sociology \cite{wang2016understanding}. The origins of ANT can be traced back to science and technology studies in the 1970s \cite{Garrety2014}. It was influenced by grounded theory and semiotics, as demonstrated in the ethnographic work of Bruno Latour and Steve Woolgar at the Salk Institute \cite{Garrety2014}.

Actor-network theory views the world as a web of interconnected relationships, where everything from people and ideas to technologies and objects plays an active role in creating the outcomes we observe. ANT enables the study of assembling and stabilizing diverse human and non-human entities within diffuse socio-material systems \cite{alcadipani2010actor}, such as blockchains.

Actor-network theory (ANT) offers valuable insights for analyzing blockchain systems. This paper leverages the following key propositions of ANT.

\begin{description}

\item [Heterogeneous Networks:] ANT posits that the social, technical, natural, and conceptual elements of the world are interconnected in heterogeneous networks \cite{wang2016understanding}. These entities are called actants, which can be both human and non-human, such as organizations, animals, technological artifacts, and concepts. Networks can be messy and inconsistent, containing contradictions and conflicts. There is no single, true representation of reality.

\item [Generalized Symmetry:] There is no distinction between "human" and "non-human" actors. Non-human actors also have their own interests. Both influence the network equally. This includes entities like animals, technologies, texts, and even natural phenomena \cite{whittle2008actor}. This approach avoids privileging certain types of entities over others in explaining social phenomena.

\item [Evolution:] Reality is never fixed but is always under construction \cite{alcadipani2010actor}. ANT focuses on how networks are constructed and how actors (entities within the network) assume roles and gain influence. It helps us understand how networks emerge, stabilize, or change over time \cite{couldry2008actor}. According to ANT, maintaining a network requires ongoing, repeated interactions and alignment of interests between actors; otherwise, it leads to the network's dissolution.

\end{description}

ANT provides an excellent theoretical lens for responsible blockchain development.

\begin{description}

\item [Network-Centric View:] At its core, blockchain is a network of interconnected nodes storing and validating data \cite{abdulkader2023efficient}. This aligns perfectly with the network-centric perspective of ANT, where actors (both human and non-human) negotiate and translate meanings to establish temporary social orders. ANT suggests that one design goal of the blockchain is to keep the network stable, ensuring that actors remain enrolled in the network without it collapsing.

\item [Stakeholder Inclusivity:] The "Generalized Symmetry" principle of ANT dismantles the human-centric view \cite{dolwick2009social} by recognizing all elements within the blockchain network as "actors," whether they are developers, users, miners, code, smart contracts, or even the underlying computational infrastructure. This holistic view is crucial for responsible development, as neglecting non-human actors can lead to unforeseen environmental, social, and economic consequences. Responsible development translates this principle into inclusive governance models, ensuring diverse voices are heard and addressed.

\item [Heterogeneity and Interdependence:] ANT emphasizes the heterogeneity of actors, acknowledging their diverse interests, values, and capabilities. Responsible development requires understanding these varied perspectives and fostering interdependence, ensuring no single actor dominates the network and decisions prioritize collective well-being.

\item [Dynamics of Alignment:] ANT highlights the dynamic nature of blockchain networks, where actors constantly negotiate and align their interests to maintain network stability \cite{wang2016understanding}. ANT encourages flexible and adaptive governance mechanisms. Responsible development translates this principle into open and transparent processes for deliberation, ensuring ongoing alignment with evolving ethical, social, and environmental considerations.

\item [Methodology:] Beyond an abstract understanding, ANT is also a methodology \cite{alcadipani2010actor}. Its methodological tools equip developers with practical frameworks for mapping the intricate relationships within the network, identifying potential power imbalances, and assessing the ethical implications of design choices.
\end{description}

\subsection{Responsible Blockchain Design Principles}

Actor Network Theory (ANT) provides a comprehensive approach to responsible blockchain design by emphasizing a network-centric perspective, stakeholder inclusivity, and the importance of maintaining a dynamic balance among diverse actors. ANT encourages the recognition of both human and non-human elements within the blockchain as critical stakeholders, promoting designs that prioritize network stability, inclusive governance, and the integration of varied interests and values. We identify the following ANT principles to be particularly helpful for the responsible development of blockchain.

\begin{itemize}
\item \textbf{S}ustainable
\item \textbf{T}ransparent
\item \textbf{E}thical
\item \textbf{A}daptive
\item \textbf{D}ecentralized
\item \textbf{I}nclusive
\end{itemize}

We call these principles the \textbf{STEADI} principles. We elaborate on each of the principles below.

\paragraph{Sustainable:}

Sustainability in the context of responsible blockchain encompasses various dimensions, including environmental \cite{arshad2023systematic}, economic, and social sustainability \cite{jiang2022tertiary}. Environmental sustainability involves reducing the energy consumption and generation of electronic waste associated with blockchain operations \cite{taylor2020blockchain}. Efforts may include adopting energy-efficient consensus mechanisms, harnessing renewables, and using efficient hardware. Economic sustainability is about creating a sustainable economic framework for the blockchain that supports its ongoing functionality and motivates involvement. Social sustainability aims to foster equal opportunities for access and engagement within the blockchain community \cite{jiang2022tertiary}.

\paragraph{Transparent:}

Transparency refers to the characteristic of blockchain systems that ensures the visibility and accessibility of information to all participants involved, without compromising privacy or security. This transparency can be achieved through various means such as openness, auditability, traceability, and explainability. Openness \cite{schmeiss2019designing} makes the blockchain's rules, governance, and data readily accessible to all participants. This fosters trust and allows for community scrutiny and participation. Auditability refers to the ability to trace and verify transactions on the blockchain easily. Every transaction is recorded in a way that is immutable and time-stamped, enabling a clear and accessible audit trail for all participants. Traceability offers a means to securely record, store, and verify the authenticity of information across a blockchain \cite{agrawal2021blockchain}. This enables accountability and helps prevent fraud and misuse.

\paragraph{Ethical:}

"Ethical" refers to the principles and practices that ensure the blockchain is developed and used in a manner that is fair, responsible, and aligned with the interests of all stakeholders involved. There are several key aspects of ethical blockchain development: fairness, accountability, privacy, and alignment of interests. Fairness in blockchain design offers equal opportunities for participation without favoritism or bias. It includes measures to resist manipulation, ensuring that the blockchain operates in a manner that is just and equitable for all users \cite{tan2023ethical}. Accountability involves establishing clear mechanisms for holding actors accountable for their actions on the blockchain, which may include dispute resolution protocols and enforcement mechanisms \cite{mishra2023height}. Privacy concerns require balancing transparency with individual privacy needs, potentially through anonymization techniques, selective data disclosure, and robust data security measures \cite{hotz2022balancing}. Moreover, ethical blockchain initiatives aim to align the interests of all stakeholders, including users, developers, and the broader society, to foster a harmonious and equitable environment.

\paragraph{Adaptive:}

"Adaptive" refers to the ability of blockchain systems or architectures to evolve and adjust in response to changing requirements, technologies, or environments. ANT emphasizes the dynamic and ever-evolving nature of networks. Networks constantly break down and regenerate. Actors constantly negotiate their roles and align their interests. Adaptability can be achieved through adaptive IT artifacts \cite{li2021two} and adaptive governance \cite{hotz2022balancing}.

\paragraph{Decentralized:}

Decentralization is a core principle that underpins the operation and management of blockchain technology. Decentralization not only refers to the decentralized IT architecture but also to the distributed governance in which power and decision-making are distributed among participants rather than concentrated in the hands of a few. This can be achieved through voting mechanisms \cite{fan2023insight}, consensus algorithms, and open collaboration structures \cite{li2021two}.

\paragraph{Inclusive:}

"Inclusive" means that the blockchain ecosystem is open, accessible, and diverse for both human and non-human actors. "Open" signifies open participation \cite{pazaitis2020breaking}. Anyone and any IT artifact should be able to participate in the blockchain ecosystem without needing permission or gatekeepers. This promotes diversity, innovation, and avoids single points of failure. "Accessible" implies that the blockchain and its applications are user-friendly and accessible to people from diverse backgrounds and technical abilities. This entails simple interfaces, clear instructions, multilingual support, and the ability for people with disabilities to use them effectively and independently \cite{lyke2023exploring}. It also means the blockchain is accessible to hardware and software from different ecosystems. Diversity promotes inclusivity and participation from underrepresented human actors or non-human actors within the blockchain ecosystem, addressing gender inequalities, economic inequalities, and geographical disparities \cite{mhlanga2023block}, and imbalances between alternative IT architectures.

\subsection{Responsible Blockchain Development Methodology}

In this section, we introduce the Actor-Network Theory-based Responsible Development Methodology (ANT-RDM). The development of blockchain refers not only to the development of hardware and software but also to the design of the network topology, consensus mechanisms, governance structures, policies, etc. Based on ANT, a blockchain is a network of interconnected actors. The development of blockchains is a process of 'creating a temporary social order, or moving from one order to another, through changes in the alignment of interests within a network' \cite{callon1984some}. In ANT, this process is referred to as translation \cite{wang2016understanding}. Translation comprises four main stages: problematisation, interessement, enrolment, and mobilisation \cite{callon1984some}. 

\begin{figure}[ht]
    \centering
    \begin{tikzpicture}[node distance=2cm, auto]
        \tikzset{node style/.style={circle, draw=black, thick, text centered, minimum size=3cm}}
        \tikzset{arrow/.style={-{Latex[length=3mm]}, thick, bend left=20}}
        \tikzset{reverse arrow/.style={-{Latex[length=3mm]}, thick, bend left=20}}

        \node[node style] (problematisation) at (0,0) {Problematization};
        \node[node style] (interessement) at (6,0) {Interessement};
        \node[node style] (enrolment) at (6,-6) {Enrolment};
        \node[node style] (mobilisation) at (0,-6) {Mobilisation};

        \draw[arrow] (problematisation) to (interessement);
        \draw[arrow] (interessement) to (enrolment);
        \draw[arrow] (enrolment) to (mobilisation);
        \draw[arrow] (mobilisation) to (problematisation);
        \draw[reverse arrow] (interessement) to (problematisation);
        \draw[reverse arrow] (enrolment) to (problematisation);
    \end{tikzpicture}
    \caption{Actor-Network-Theory-based Responsible Blockchain Development Methodology}
    \label{fig:ant-rbdm}
\end{figure}
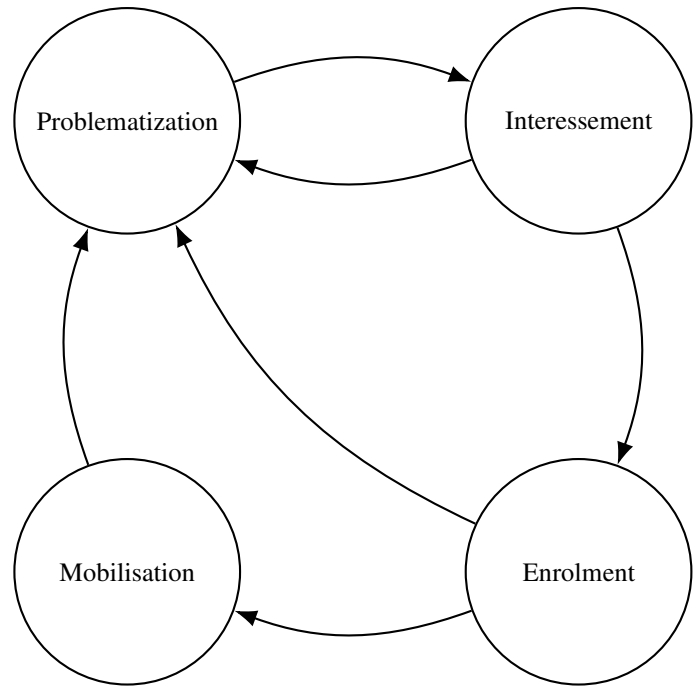

\subsubsection{Problematisation}

The problematisation stage sets the foundation for the entire translation process. It is where you define the issue, frame it in a compelling way, and gather the key actors needed to address it. Here are four key actions to focus on during this stage:

\paragraph{Define the problem}

Clearly identify the specific issue or challenge that blockchain technology could potentially address. Pinpoint the core pain point and its underlying causes. The problems may be external, such as a lack of trust and societal inefficiencies, or internal to the blockchain itself, including issues like scalability, security, and energy consumption.

\paragraph{Identify "focal actors"}

A focal actor is an entity within a network that holds significant influence over other actors and the overall network dynamics \cite{wang2016understanding}. They actively shape the network through their actions. They drive the process of translation and gather other actors’ support.

The focal actors can be identified by the following factors:

1. Resources: Focal actors control valuable resources like funding, information, or expertise, which give them leverage over others in the network. Think of them as the ones powering the engine of the project. For example, a company providing crucial tech infrastructure or a venture capitalist with substantial funding would be resource-rich focal actors.

2. Knowledge: A deep understanding of blockchain technology, specific market niches, or the project's technical complexities makes certain actors critical problem-solvers and decision-makers. Their expertise becomes indispensable for the project's progress. Imagine a team of seasoned blockchain developers or a regulatory expert guiding the way through legal hurdles – these are knowledge-rich focal actors.

3. Relationships: Strong connections and alliances with other key players further solidify a focal actor's position and amplify their influence. Think of them as network builders and facilitators. For example, a well-connected industry consortium or a partnership with a reputable platform can open doors and bring diverse stakeholders together – these are relationship-rich focal actors.
By identifying and engaging with the focal actors, a blockchain project can leverage their combined resources, knowledge, and network to gain crucial support and funding, navigate complex technical challenges, build trust and legitimacy within the broader ecosystem, and reach a wider audience and secure user adoption.

\paragraph{Identify all relevant actors}

The goal of responsible blockchain development is to develop blockchain systems that prioritize ethics, sustainability, and inclusivity. By including a wide range of actors, the development of responsible blockchains takes into account the needs and concerns of all stakeholders. This can help to avoid unintended consequences and ensure that blockchain technology is used in a way that benefits everyone. Who are the stakeholders most affected by the problem? Who else would need to be involved in the network to develop and implement the blockchain solution? Potential allies and opponents, existing power dynamics, and any potential conflicts of interest within the network should be considered at this step.

From an ANT perspective, blockchain ecosystems are dynamic networks composed of a diverse range of actors, both human and non-human:

\textbf{Human Actors} may include consumers, who use blockchain-based applications for activities like financial transactions, supply chain management, or voting. Miners and validators play a crucial role in creating and verifying new blocks on the blockchain, often receiving cryptocurrency or fees as rewards. Full nodes, maintained by individuals or organizations, hold a complete copy of the blockchain ledger, aiding in network security and decentralization. The category of software developers encompasses those who develop the software itself, create smart contracts, and develop blockchain protocols.

Infrastructure providers in the blockchain space include node hosting services, which provide the necessary infrastructure to run full nodes and support network operations. Mining pools are collaborations among miners, pooling computing resources to enhance their chances of finding blocks and earning rewards. Wallet providers develop both software and hardware solutions that enable users to store their digital assets and interact with the blockchain. Cryptocurrency exchanges provide the platforms where users can trade cryptocurrencies and blockchain derivatives.

Regulators and policymakers in the blockchain sector involve government agencies that develop and enforce blockchain-related regulations, influencing adoption and use cases. Standard-setting organizations define industry standards and best practices for blockchain development and implementation. Self-regulatory organizations set voluntary guidelines and compliance programs for actors in specific blockchain ecosystems.

Researchers and academics conduct research on various aspects of blockchain technology, contributing to its future development and applications. They may include educators, trainers, futurists, and visionaries.

\textbf{Non-Human Actors} may include the Blockchain Protocol, which acts as the foundational technological infrastructure, enabling interactions and value exchanges. Smart Contracts, as self-executing code on the blockchain, shape interactions and facilitate transactions. Cryptocurrencies and Tokens, as digital assets inherent to the blockchain ecosystem, function as mediums of exchange and stores of value. Mining Hardware and Infrastructure represent the physical technology necessary for mining and maintaining the network. Lastly, Software Tools and Applications provide support for development, wallet management, and interaction with the blockchain system.

\paragraph{Define the OPP}

The focal actor needs to establish an obligatory passage point (OPP), which refers to a situation or process specified by the focal actor through which relevant actors can achieve a shared interest \cite{sarker2006understanding}. The focal actor should analyze the current state of the network, existing challenges, and limitations in the context where blockchain is proposed to be implemented. This relies on the previous step "Define the problem." Then, the focal actors define the OPP and explain why the network needs to pass through the proposed OPP to achieve their interests.

\subsubsection{Interessement}

Interessement is about 'interesting' or engaging the actors in the network. At this stage, focal actors convince other actors to accept the OPP defined in the previous stage. By this stage, the focal actors have identified all the relevant actors and have a good understanding of each of their interests. Focal actors should actively involve them in the project, aligning their interests with the goals of the blockchain. This process often involves negotiation among actors, and incentives may be provided so that other actors are willing to pass through the OPP \cite{wang2016understanding}. The focal actors can use demonstrations, prototypes, pilot projects, simulations, or theoretical arguments to convince other actors that the proposed network will create a "better" social order compared to existing alternatives. During this stage, it's also crucial to address any conflicts or competing interests among actors. This could involve negotiating terms, modifying aspects of the blockchain to suit different needs, or even going back to the previous stage and redefining the problems and OPP.

\subsubsection{Enrolment}

Enrolment is where the actual development and implementation of the blockchain take place. This step is critical because it transforms the concept into materialization and acceptance. It is important to note that the development of responsible blockchains is not the mere creation of IT artifacts; it is the creation of a new social order, a new network that may stabilize and sustain. They may include:

\paragraph{IT Artifacts:} To develop the hardware, software, IT infrastructure, network architecture, security protocols, and any other necessary technical elements based on the specific application and actors' needs.

\paragraph{Blockchain Governance:} Governance in blockchain refers to the mechanisms, policies, and procedures that determine how decisions are made within a blockchain network. Effective governance is crucial for the sustainability, adaptability, and trustworthiness of blockchain systems. Issues such as decision-making processes, consensus mechanisms, transparency, accountability, inclusiveness, representation, fork management, regulatory compliance, smart contract governance, upgradability, adaptability, economic incentives, and penalties should be considered.

\paragraph{Regulatory Frameworks:} Clear and supportive regulatory guidelines are crucial for the adoption and integration of blockchain technology. This involves developing laws and regulations that address issues such as private property, intellectual property, data privacy, security, financial transactions, and cross-border legal implications.

\paragraph{Standardization and Interoperability:} Establishing industry standards to ensure interoperability between different blockchain systems. This includes technical standards for data formats, protocols, and interfaces, as well as operational standards for governance, auditing, and compliance. Standardization can enhance the scalability and integration of blockchain systems across various industries.

\paragraph{Protocols:} Developing protocols for blockchains involves creating a set of rules and standards to govern the operation, security, interoperability, scalability, and sustainability of blockchain networks. The protocols may include consensus protocols, security protocols, scalability protocols, smart contract protocols, governance protocols, and data storage and management protocols. These protocols are crucial for ensuring the efficiency, trustworthiness, and broader adoption of blockchain technology.

\paragraph{Education and Training:} Developing educational resources and training programs to increase blockchain literacy among developers, users, and stakeholders is vital. This includes not only technical training but also education about the legal, ethical, and business aspects of blockchain.

\paragraph{Ethical and Social Frameworks:} Addressing the ethical and social implications of blockchain technology is important. This includes considering issues of fairness, privacy, digital divide, and the potential societal impacts of widespread blockchain adoption.

\paragraph{Economic Models:} Creating incentives and reward systems. Developing incentive mechanisms to encourage ongoing participation and contribution is another critical aspect of enrolment. This could involve financial incentives, recognition, or other benefits that motivate actors to remain actively involved in the blockchain network.

\paragraph{Ecosystem Development:} Building a supportive ecosystem around blockchain technology is crucial. This involves fostering collaborations between startups, established companies, governments, educational institutions, and other stakeholders. A healthy ecosystem can spur innovation, provide funding opportunities, and facilitate knowledge exchange.

In the end of enrolment stage, non-human actors such as the IT artifacts and governance policies are developed, and their interests are inscribed by their developers \cite{wang2016understanding}. All the human and non-human actors are enrolled and the new social order and network are adopted.  

\subsubsection{Mobilisation}

Mobilization in blockchain development represents the final stage of translation, where a temporary social order solidifies around the block-chain, achieving stability through continuous adoption, usage, and maintenance. This process involves uniting various actors and resources into a stable network dedicated to maintaining and utilizing the blockchain. Activities may include:

\paragraph{Ensuring Representation of Interests:} In the mobilisation stage, it is crucial that the interests and entities that have been enrolled in the earlier stages are adequately represented. This means that the blockchain technology and governance must effectively embody the needs, expectations, and desires of the various actors involved, such as developers, users, investors, and regulatory bodies.

\paragraph{Facilitate Ongoing Translation and Negotiation:} Continuously adapt the network in response to changing needs and feedback from actors. Foster ongoing collaboration and participation.

\paragraph{Monitor and Address Power Dynamics:} Be aware of the potential for new inequalities to emerge within the network. Work to maintain a balance of power and prevent any single actor from gaining undue control.

\section{Future Research Agenda}
Actor-Network Theory-based Responsible Development Methodology (ANT-RDM) is a novel approach to system development. However, its effectiveness has yet to be fully tested across diverse contexts. There is also a need to develop supporting tools, metrics, and educational programs to facilitate broader adoption of this methodology.

\paragraph{Empirical Validation}
Comparative studies are necessary to empirically evaluate the effectiveness of ANT-RDM across different settings. This could involve case studies, pilot implementations, controlled experiments, and longitudinal studies comparing ANT-RDM with existing methodologies such as the Waterfall Model \cite{ruparelia2010software}, Rapid Application Development (RAD) \cite{martin1991rapid}, Agile Development \cite{martin2003agile}, and DevOps \cite{erich2014report}. Such research would provide a robust empirical basis for evaluating the methodology's effectiveness.

\paragraph{Generalization}
ANT-RDM and STEADI principles are methodology and principles applicable to any system development. This paper discusses these principles within the context of blockchain development; future research could extend their application to other types of systems, including IT systems (e.g., artificial intelligence systems), and non-IT systems (e.g., organizational or social structures). Empirical studies should also test ANT-RDM and its principles across various cultural contexts.

\paragraph{Performance Metrics}
From the ANT-RDM perspective, system performance encompasses more than just the effectiveness and efficiency of the IT artifact—it also measures how well the system aligns with and serves the interests of all stakeholders. Future research should develop metrics to assess this alignment and the perceived mutual benefits among stakeholders. Metrics should be developed to evaluate how well the system adheres to principles like sustainability, transparency, ethics, adaptability, decentralization, and inclusivity. Both primary methods (e.g., survey questionnaires) and secondary measures need to be developed to monitor the health of the system effectively.

\paragraph{Tool Development}
Computer-aided tools are needed to support each stage of the ANT-RDM process, from problematization and interessement to enrollment and mobilization. Future developments could include dashboards for metrics tracking and collaboration tools optimized for the workflow of the methodology.

\paragraph{Educational and Training Programs}
Future research also needs to develop training programs or educational courses to help practitioners comprehend and effectively implement ANT-RDM and STEADI principles. Training could also cover the use of software tools, templates, and best practices specific to this methodology.

\section{Conclusion}

This paper provides a comprehensive analysis of the challenges and controversies of blockchain technology. It identifies the technical challenges such as scalability, security, privacy, and interoperability, as well as the business, adoption challenges, and social, economic, ethical, and environmental controversies within current blockchain systems. We argue that responsible blockchain development is key to overcoming these challenges and achieving mass adoption. This paper introduces the STEADI principles (sustainable, transparent, ethical, adaptive, decentralized, and inclusive) for responsible blockchain development. Based on Actor-Network Theory, this paper also introduces the Responsible Blockchain Development Methodology (RBDM), which includes the steps of problematisation, interessement, enrolment, and mobilisation. This methodology emphasizes the interplay between human actors (developers, users, regulators) and non-human actors (technology, protocols, code, and environment) and encourages the constant alignment of diverse interests within the network to keep the blockchain vibrant and stable.

\bibliographystyle{unsrt}  
\bibliography{references}  

\end{document}